\documentclass[useAMS,usenatbib]{mn2e}

\usepackage{graphicx}
\usepackage{txfonts}

\begin{document}

\title[Emission from Sagittarius A*]{A polarised infrared flare from Sagittarius A* and the signatures of orbiting plasma hotspots\thanks{Partially based on observations at the Very Large Telescope (VLT) of the European Southern Observatory (ESO), Chile.}}

\author[S. Trippe et al.]
{
S. Trippe,$^1$\thanks{E-mail: {\tt trippe@mpe.mpg.de}} T. Paumard,$^1$\thanks{Present adress: Observatoire de Paris - Section de Meudon, 5 Place Jules Janssen, F-92195 Meudon Cedex, France}  T. Ott,$^1$ S. Gillessen,$^1$ F. Eisenhauer,$^1$ F. Martins$^1$ and R. Genzel$^{1,2}$
\\
\\
$^1$Max-Planck-Institut f\"ur extraterrestrische Physik, Postfach 1312, D-85741 Garching, Germany \\
$^2$Department of Physics, University of California, CA 94720, Berkeley, USA
}

\date{ Accepted 2006 November 23. Received 2006 November 22; in original form 2006 September 22 }

\pagerange{\pageref{firstpage}--\pageref{lastpage}} \pubyear{2006}

\maketitle

\label{firstpage}

\begin{abstract}
In this article we summarise and discuss the infrared, radio, and X-ray emission from the supermassive black hole in the Galactic Centre, SgrA*. We include new results from near-infrared polarimetric imaging observations obtained on May 31st, 2006. In that night, a strong flare in $K_s$ band (2.08 $\mu$m) reaching top fluxes of $\sim$16 mJy could be observed. This flare was highly polarised  (up to $\sim$40 \%) and showed clear sub-structure on a time scale of 15 minutes, including a swing in the polarisation angle of about 70 degrees. For the first time we were able to observe both polarised flux and short-time variability, with high significance in the same flare event. This result adds decisive information to the puzzle of the SgrA* activity. The observed polarisation angle during the flare peak is the same as observed in two events in 2004 and 2005. Our observations strongly support the dynamical emission model of a decaying plasma hotspot orbiting SgrA* on a relativistic orbit. The observed polarisation parameters and their variability with time might allow to constrain the orientation of accretion disc and spin axis with respect to the Galaxy.
\end{abstract}


\section{Introduction}

The centre of our Milky Way hosts the 3.6-million-$M_{\odot}$ supermassive black hole and radio source SgrA*. This black hole is generally invisible in NIR wavelengths and was not detected in this spectral range before 2002 when diffraction-limited observations at 8-m-class telescopes became possible (Genzel et al. \cite{genzel2003}, Ghez et al. \cite{ghez2004}).

Since then, several NIR flares, which appear on time scales of few events per day,  have been observed photometrically (Ghez et al. \cite{ghez2005}, Eckart et al. \cite{eckart2006a}), spectroscopically (Eisenhauer et al. \cite{eisenhauer2005}), and polarimetrically (Eckart et al. \cite{eckart2006b}). Such flares last typically for about 60--120 minutes. These observations gave information on the colours and spectral indices of flares (Ghez et al. \cite{ghez2005}, Eisenhauer et al. \cite{eisenhauer2005}, Gillessen et al. \cite{gillessen2006}, Krabbe et al. \cite{krabbe2006}). They included the detection of polarised flux and quasi-periodic substructures on time scales of 15 to 20 minutes (Genzel et al. \cite{genzel2003}, Eckart et al. \cite{eckart2006b}).

In recent years, variable and flaring emission from SgrA* has been observed in a variety of wavelength bands, especially in the radio (Aitken et al. \cite{aitken2000}, Melia \& Falcke \cite{melia2001} [and references therein], Bower et al. \cite{bower1999a, bower1999b, bower2003a}, Miyazaki et al. \cite{miyazaki2004}, Marrone et al. \cite{marrone2006}) and X-ray (Baganoff et al. \cite{baganoff2001, baganoff2003}, Goldwurm et al. \cite{goldwurm2003}, Aschenbach et al. \cite{aschenbach2004}, B\'elanger et al. \cite{belanger2005, belanger2006}) regimes. Sub-structure on minute time scales was also detected in X-ray flares (Aschenbach et al. \cite{aschenbach2004}, B\'elanger et al. \cite{belanger2006}). Additionally, variable polarised flux from SgrA* was found in sub-mm to mm radio bands (Bower et al. \cite{bower2005}, Marrone et al. \cite{marrone2006}, Macquart et al. \cite{macquart2006}).

In this article we discuss the physics behind the emission from SgrA* taking into account new results of polarimetric imaging observations obtained in May 2006. In section 2 we describe the observations and the data reduction, in section 3 we present the observational results. In section 4 these data are placed into the context of earlier results, and in section 5 they are discussed and interpreted.

\section{Observations and data reduction}

We have repeatedly carried out observations on the 8.2-m-UT4 (Yepun) of the ESO-VLT on Cerro Paranal, Chile, using the detector system NAOS/CONICA (NACO for short) consisting of the AO system NAOS (Rousset et al. \cite{rousset2003}) and the 1024$\times$1024-pixel NIR camera CONICA (Lenzen et al. \cite{lenzen2003}).

\begin{figure}
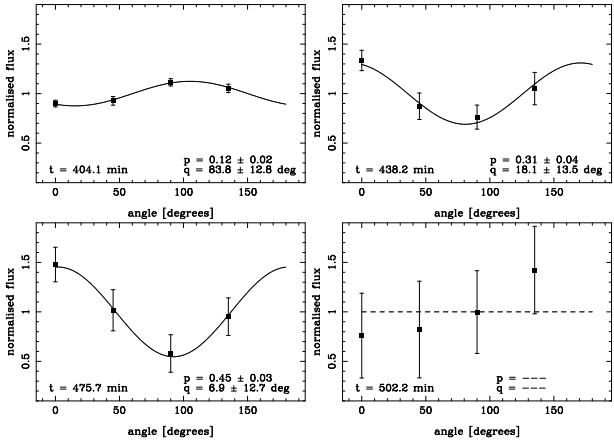


\centering

\includegraphics[height=4cm,angle=-90]{fig01a.eps}
\includegraphics[height=4cm,angle=-90]{fig01b.eps}

\includegraphics[height=4cm,angle=-90]{fig01c.eps}
\includegraphics[height=4cm,angle=-90]{fig01d.eps}

\caption{Four examples for the calculation of degrees and angles of polarisation as presented in Figs. \ref{fluxes} and \ref{params}. Each panel shows the normalised and calibrated flux of SgrA* in the four polarimetric channels belonging to a given time bin. The four flux values are plotted together with the best fitting sine curve used to compute the degree of polarisation $p$ and the polarisation angle $q$ at time $t$. In each case, $p$ is given by the amplitude of the sine and $q$ by the phase. In cases where the flux values are identical within the errors (bottom right panel), no polarisation parameters are calculated.}

\label{channels}

\end{figure}


\begin{figure}
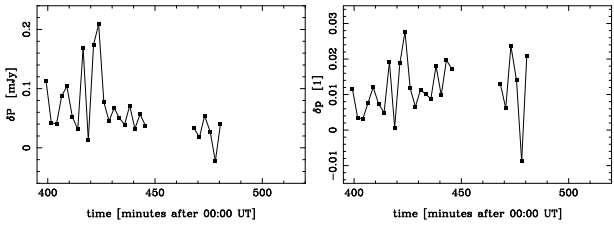


\centering

\includegraphics[height=4cm,angle=-90]{fig02a.eps}
\includegraphics[height=4cm,angle=-90]{fig02b.eps}

\caption{Comparison of two methods for calculating polarisation parameters. Polarised flux $P$ and polarisation fraction $p$ were computed (a) numerically via a sine fitting method, (b) analytically using the Stokes parameters $I$, $Q$, and $U$. The analytical approach was explicitely corrected for bias. Both panels show the differences between the two approaches; this allows to estimate the influence of bias. {\sl Left panel}: Differences in polarised flux. The average deviation betwen the two methods is 0.064$\pm$0.073~mJy. {\sl Right panel}: Differences in polarisation fraction. The average deviation is 0.012$\pm$0.013. As all differences are zero within the quoted errors (see also Figs. \ref{fluxes}, \ref{params}), the bias is not significant.}

\label{bias}

\end{figure}

On May 31st, 2006, in total 240 minutes of polarimetric $K_s$ band ($\lambda_{\rm center} = 2.08 \mu$m) imaging data of the Galactic Centre were obtained. The Wollaston prism mode of NACO made it possible to simultaneously observe two orthogonal polarisation angles (corresponding to the ordinary and the extraordinary beam of the prism respectively) per image. In order to cover a sufficient number of polarimetric channels, the observed angles were switched using a half-wave retarder plate. 

The images were obtained alternately covering the polarisation angles 0$^{\circ}$ / 90$^{\circ}$ and 45$^{\circ}$ / 135$^{\circ}$ respectively. Each cycle took no more than about 150 seconds. The spatial resolution of the data is around 60 mas at mediocre Strehl ratios. All frames have a pixel scale of 13.27 mas/pixel.

All images were sky subtracted, bad-pixel- and flat-field-corrected. To extract the fluxes of SgrA* and two comparison stars, we applied aperture photometry. 8 bright and isolated stars in the field of view served as calibrator sources. As SgrA* was confused with a weak star, S17, at the observation epoch, this star's flux contribution of 2.5 mJy was subtracted from the flare data.

As always two pairs of polarisation channels were observed alternately, for each source sets of four flux values (for the four angles) per time bin were obtained. In order to extract the polarimetric parameters -- degree of polarisation $p$, angle of polarisation $q$ -- from a given data set, this data set was first normalised by dividing all four values by their average. Due to this, the average of the data set is set to 1. The amplitude of variations \emph{around the average level} due to polarised flux is limited to the range from 0 to 1. As we use the convention that the degree of polarisation is the ratio of polarised flux vs. total flux, this amplitude corresponds to the degree of polarisation.  We compute the polarised flux as the product of the degree of polarisation and the total source flux. One should note that throughout this article ``polarisation'' and ``polarimetry'' refer to {\it linear} polarisation.

\begin{figure}
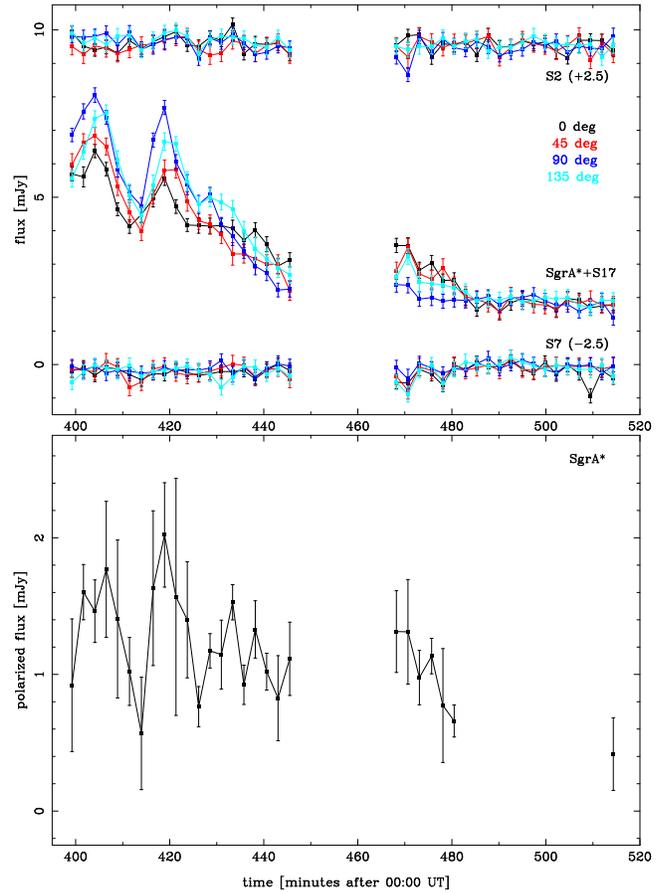


\centering

\includegraphics[height=8.5cm,angle=-90]{fig03a.eps}

\includegraphics[height=8.5cm,angle=-90]{fig03b.eps}

\caption{Observed total and polarised fluxes of SgrA* and the comparison stars S2 and S7. {\sl Top panel:} Lightcurves of SgrA*, S2, and S7 separated into the four polarimetric channels. Due to the splitting of the light into one ordinary and one extraordinary beam, each channel in average contains one half of the total source flux. These values are the observed fluxes, before calibrating relative to the extinction screen polarisation and subtraction of the confusing star S17. S2 and S7 are shifted along the flux axis. The fast variability of SgrA* and its strong polarisation can be seen clearly, especially in comparison to S2 and S7. {\sl Bottom panel:} polarised flux of SgrA*. The two main maxima in the polarised flux correspond to the maxima in the lightcurve. In all figures the central gap in the data is due to sky observations, around a time $t=480$ min the flare fades out. Where data points are missing, no parameters were computed (see Sect. 2 for details).}

\label{fluxes}

\end{figure}

As described above, in each image source fluxes are calibrated by dividing the target source count rates by the count rates of calibration stars taken from the same image (photometric calibration). This means that a possible average polarisation of the calibrator stars is erased. As indeed the stars of the observed Galactic Centre region show an average polarisation  of 4\% at an angle of 25$^{\circ}$ due to foreground extinction by dust (Eckart et al. \cite{eckart1995}, Ott et al. \cite{ott99}), this has to be corrected. This correction was done by multiplying the four flux values of a data set with the function

\begin{equation}
f(\phi) = 1 + 0.04\cdot\sin(2\cdot(\phi + 25^{\circ})) \ , \ \ \phi = 0^{\circ}, 45^{\circ}, 90^{\circ}, 135^{\circ}
\end{equation}

\noindent
(polarimetric calibration). The factor 2 in the argument of the sine is due to the convention that polarisation angles are limited to the range [0$^{\circ}$, 180$^{\circ}$]. After this, each normalised data set was fit with a sine curve with a period of 180$^{\circ}$. This delivers the degree of polarisation (the sine curve's amplitude) and the angle of polarisation (the sine curve's phase).

In our definition, a polarisation angle of 0$^{\circ}$ corresponds to a pointing to the north and the angle is counted east of north. In both the polarimetric calibration and the fitting, a global rotation of the polarisation vector with respect to the sky of $36^{\circ}$ was taken into account. This rotation was found using polarimetric NACO Wollaston data of the calibrator star IRS21 obtained in July 2005. Comparing the parameters extracted from this data set with results found with different instruments and reported earlier (14\% and 14$^{\circ}$; Eckart et al. \cite{eckart1995}, Ott et al. \cite{ott99}) leads to a rotation of $34^{\circ}$. This rotation might be caused by a known shift (the abovementioned $36^{\circ}$) in the zero position of the half-wave plate, although this shift is assumed to be corrected in the instrument setup (N. Ageorges, {\it priv. comm.}; NACO HWP commissioning report). Given the numerical agreement (better than $2^{\circ}$) we believe that at least in this data set a correction was not applied.

The sine fitting procedure described above is demonstrated in Fig. \ref{channels} for four different time bins. Here the connection between $p$ and $q$ on the one side and amplitude and phase of the flux data on the other side is obvious. This figure already indicates some evolution of the polarisation parameters with time; details will be discussed below.

\begin{figure}
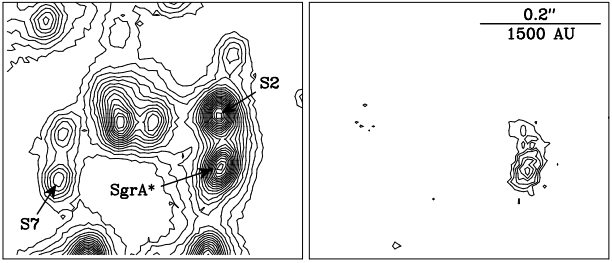


\centering

\includegraphics[height=4cm,angle=-90]{fig04a.eps}
\includegraphics[height=4cm,angle=-90]{fig04b.eps}

\caption{Contour maps of integrated and  polarised emission from SgrA*; N is up, E to the left. {\sl Left:} Sum image of the channels 0$^{\circ}$ and 90$^{\circ}$ at the time of the first flux peak ($t=405$) showing the immediate vicinity of SgrA*. Contours are 20, 25, ... 120 times noise level. {\sl Right:} Difference of the channel maps used for the left image. Contours are 4, 6, ... 16 times noise level. The strong residual source at the position of the flare -- corresponding to the polarised flux  -- is clearly visible.}

\label{photos}

\end{figure}

Data sets, for which the values were consistent with being identical within the errors (corresponding to a $\chi^2_{\rm reduced}<0.789$ in case of three degrees of freedom) were not fit in order not to apply a systematically incorrect model. Another effect to be taken into account was the bias caused by the non-zero errors of the flux values. Bias can lead to a systematic overestimation of polarisation fraction and polarised flux especially in cases of low fluxes. In order to check this, we calculated $p$ and polarised flux using analytical expressions for the Stokes parameters $I$, $Q$, and $U$ corrected for the bias terms. The results are presented in Fig. \ref{bias}. In no case the deviation exceeded 3\% (in $p$; average deviation: 1.15$\pm$1.34\%) resp. 0.2~mJy (polarised flux; average deviation: 0.064$\pm$0.073~mJy). As all deviations are zero within the quoted errors, the influence of bias can be safely neglected for the further discussion.

\section{Results}

The lightcurves for SgrA* and two comparison stars, S2 and S7, are presented in Fig. \ref{fluxes} (top panel). In this figure the lightcurves are shown for each polarisation channel separately. The presented values are the observed fluxes before polarimetric calibration relative to the foreground extinction screen and before subtracting the flux of S17. In all lightcurves gaps due to intermediate sky observations are present. All times mentioned here and elsewhere in this article are minutes after 00:00 UT of the observation day, the first data point obtained is located at $t=399$ min.

\begin{figure}
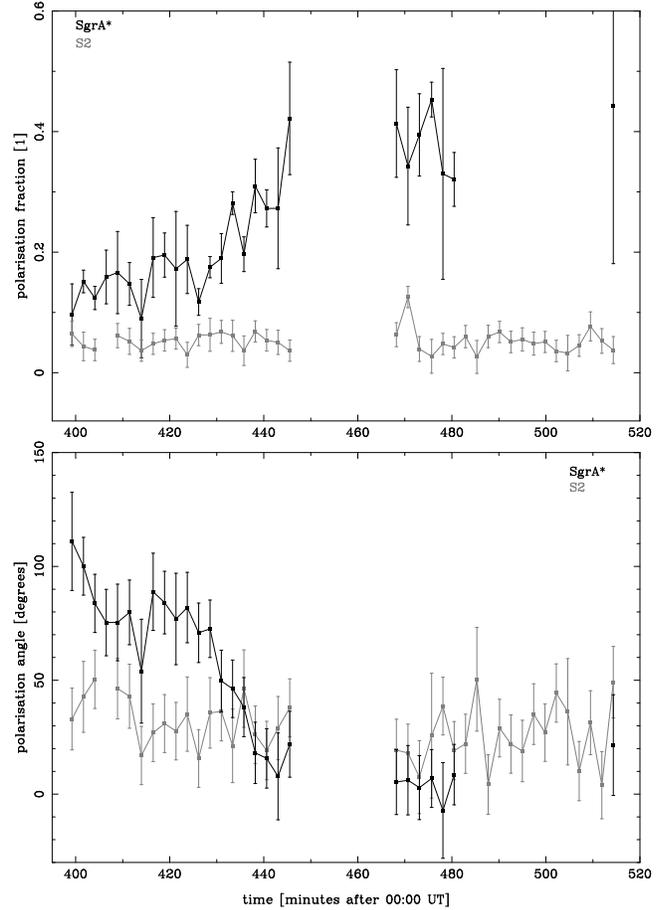


\centering

\includegraphics[height=8.5cm,angle=-90]{fig05a.eps}

\includegraphics[height=8.5cm,angle=-90]{fig05b.eps}

\caption{Evolution of degree and angle of polarisation for SgrA* and S2 during the flare. {\sl Top panel:} Degree of polarisation. During the flare ($t<480$) it never sinks below 10\%, reaching top levels of $\sim$40\% . {\sl Bottom panel:} Angle of polarisation. Of special interest is the strong swing of $\sim$70$^{\circ}$ occuring in the time range 430...445 min, i.e. within 15 minutes. In both panels the results for S2 mirror the calibration. Where data points are missing, no parameters were computed (see Sect. 2 for details).}

\label{params}

\end{figure}

From these lightcurves one can see a strong and fast variability of the flare. Especially interesting is its double peak, which shows flux variations in the order of 7 mJy or 40\% within 10 minutes time. The two maxima of the double peak are separate in time by only 15 minutes. Additionally, the different fluxes in the polarimetric channels point towards significant polarisation. This is demonstrated more directly in Fig. \ref{photos}, where the difference of two channels is mapped and compared to a sum image of the vicinity of SgrA*. It is important to note that those differences in the polarimetric channels are not visible in the lightcurves of the comparison stars. This holds for both flux levels corresponding to the peaks of the flare (S2) and for fluxes corresponding to the end of the flare (S7).

The amount of observed polarised flux from SgrA* is shown in the bottom panel of Fig. \ref{fluxes}. The polarised flux shows two maxima at times 405 resp. 420 reaching up to around 2 mJy; these maxima correspond to the double peak in the lightcurves. During the entire flare the polarised flux is never lower than $\sim$0.6 mJy.

The evolution of the parameters degree and angle of polarisation is presented in Fig. \ref{params}. The degree of polarisation is about 15\% at the beginning of the observation and increases after the second main maximum ($t>430$) up to roughly 40\%. This can be seen also in Fig. \ref{channels}, were two examples for the fluxes in the four polarimetric channels are given for time bins corresponding to the first peak ($t=404$) resp. the end of the flare ($t=476$). The increase is a direct consequence of the fact that the polarised flux level remains roughly constant while the overall flux is decreasing with time after the double peak.

Inspecting Fig. \ref{fluxes} and \ref{params}, it might not be obvious that for times $t<430$ the polarised flux is variable and the degree of polarisation is not. To check this, we computed the reduced $\chi^2$ for each data set using the assumption that the data do not vary with time. We find a $\chi^2_{\rm reduced} = 2.12$ for the polarised flux, whereas for the degree of polarisation we find $\chi^2_{\rm reduced} = 0.88$. Thus we conclude that indeed only the polarised flux varies for $t<430$, whereas the degree of polarisation remains constant within the errors.

The polarisation angle remains at a constant level around 80$^{\circ}$ during the time of the double peak, i.e. $t\le 430$. The first two data points indicate that the angle could have been even larger before the first peak; but as this signal is only marginally significant (1--1.5$\sigma$), this is completely speculative. Beginning at $t\sim 430$, $q$ swings by about 70$^{\circ}$ within 15 minutes, reaching values down to $\sim$10$^{\circ}$. This is (within the errors) close to the angle of the foreground polarisation ($\sim$25$^{\circ}$).

When discussing the angle of polarisation, one has to keep in mind possible calibration artefacts. As described in Sect. 2, our data are calibrated so that an intrinsically unpolarised source shows a signal corresponding to $p = 4$\% and $q = 25^{\circ}$. Therefore it is in general possible to observe a source composed of two superposed flux components, a polarised one, and a non-polarised one. As long as both components are significantly bright, one would measure the polarisation parameters of the polarised source flux. But when the polarised component fades away, the observed $p$ and $q$ would move towards the foreground level.

\begin{figure}

\centering

\includegraphics[height=8.5cm,angle=-90]{fig06a.eps}

\includegraphics[width=4.1cm]{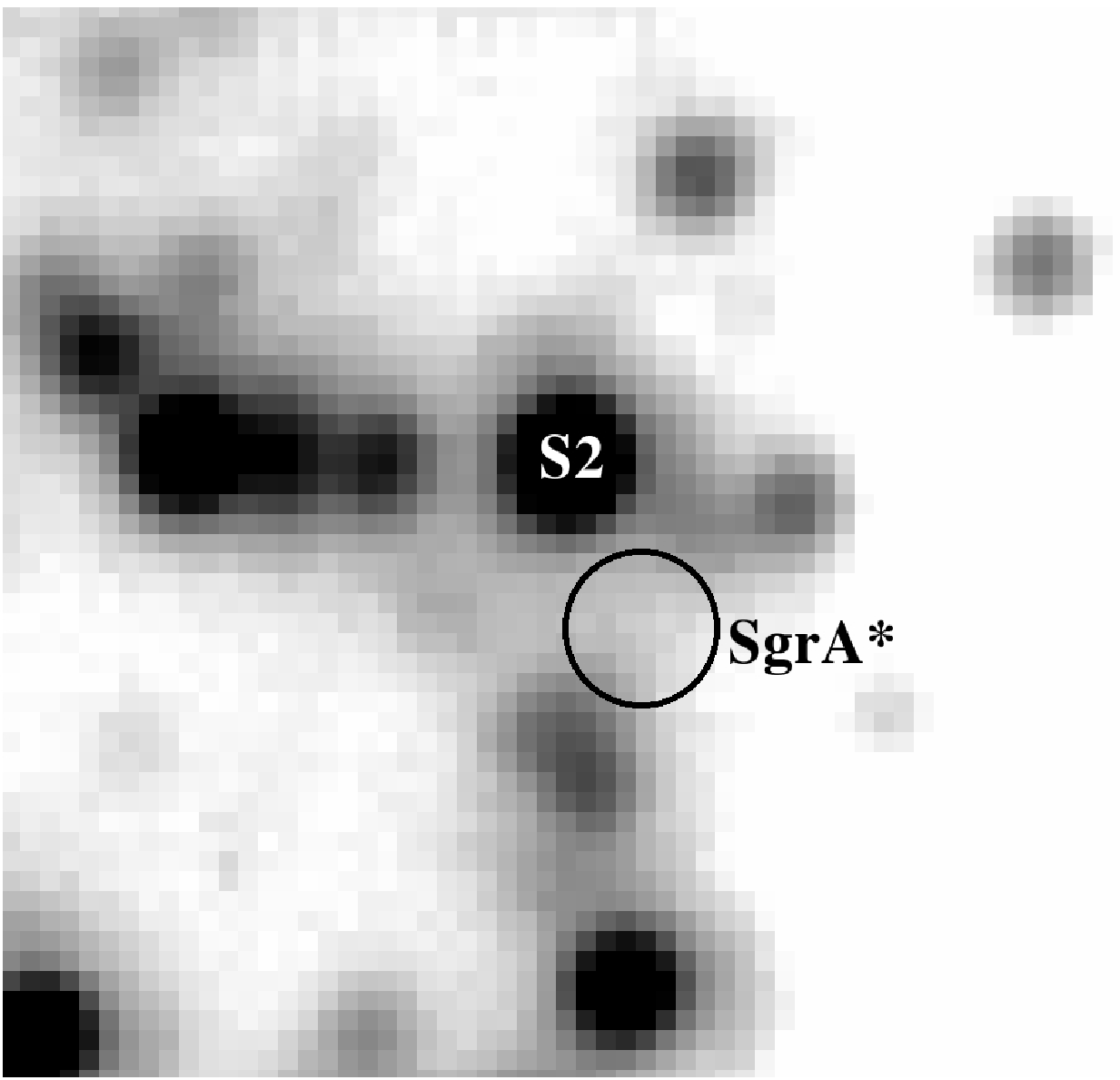}
\includegraphics[width=4.1cm]{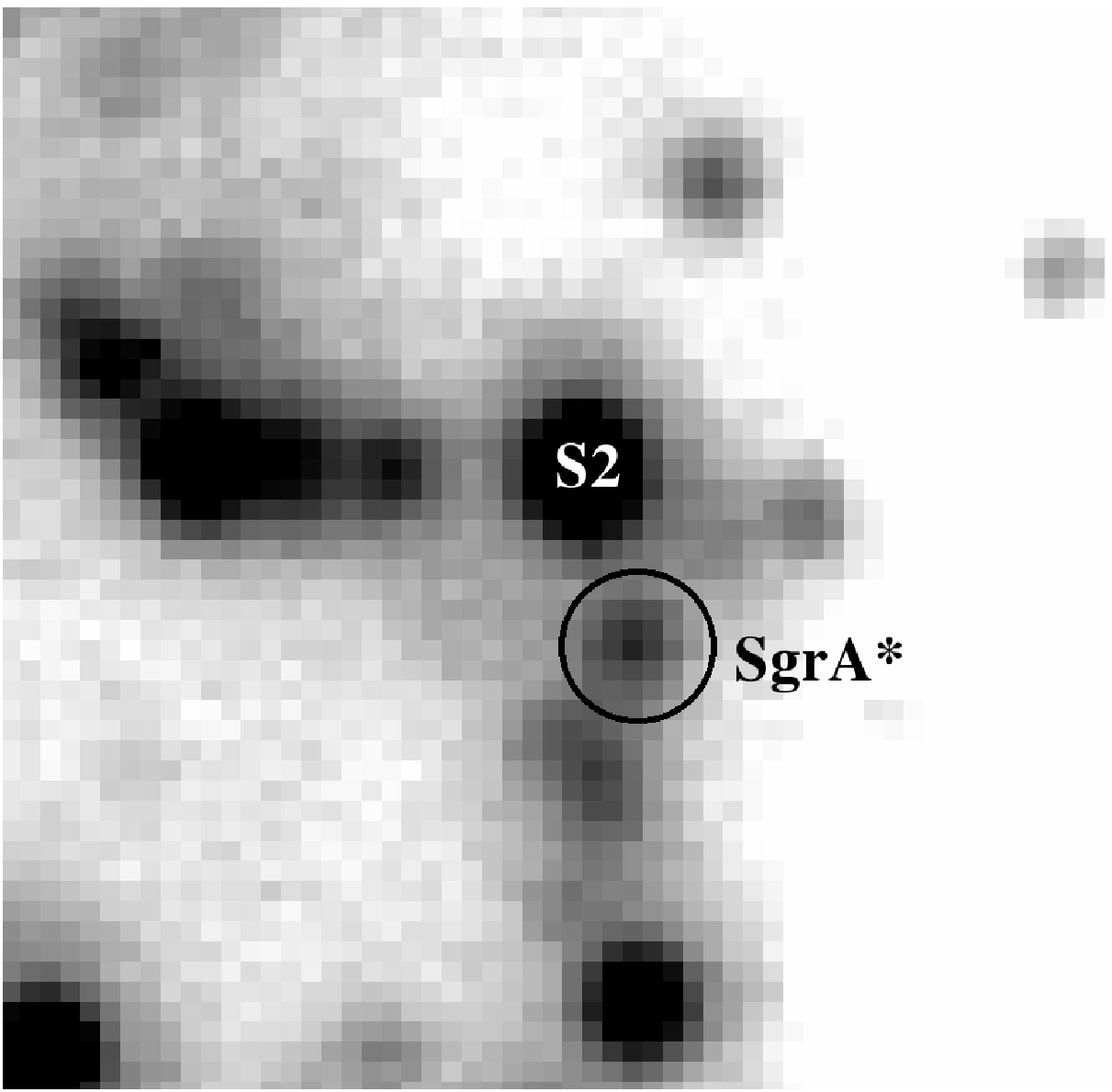}

\caption{A beginning $H$ band flare observed on April 28, 2004, with NACO. {\sl Top panel}: Lightcurve of the event. The flare begins after $t\approx$~85~min. Before this time, no flux is detected at the position of SgrA*. For $t <$~85~min the upper limits for source flux are given. Gaps are due to sky observations and a short breakdown of the AO system. {\sl Bottom panels}: Images showing SgrA* before and after beginning of the flare. The {\sl left} image is an average of 20 frames obtained in the time range $t=$45...60~min. The position of SgrA* is free of any excess emission. The {\sl right} image is an average of the last five frames obtained. Here SgrA* is clearly visible. This example data set illustrates the flaring character of SgrA*: the black hole does not show any detectable activity for at least 1.5 hours, then a flare raises from zero level within minutes.}

\label{startingflare}

\end{figure}

To assure that we are not mislead by such an effect, the comparison to a source with a brightness similar to SgrA* (plus S17) at the very end of the flare ($t=470...480$) without applying any polarimetric calibration becomes important. Indeed this is what is shown in Fig. \ref{fluxes}, where the photometric lightcurves of SgrA* and S7 are compared. As one can see, even at the very end of the flare SgrA* is clearly polarised intrinsically, whereas S7 is not at all. Additionally, we repeated the fitting of the polarisation parameters without re-introducing the foreground polarisation into the data. The results turned out to be identical within the erors compared to those shown in Figs. \ref{fluxes} and \ref{params}. Thus we are confident that the description above is valid.

Morphologically, the flare shows two phases.
In the first phase, covering times $t\le 430$, the double peak occurs, the polarised flux changes rapidly and traces the overall emission, and both degree and angle of polarisation remain constant.

In the second phase ($t>430$) the overall flare slowly fades away while the polarised flux remains on a roughly constant level, leading to an increase in the degree of polarisation. Additionally the swing in polarisation angle occurs.

Following the evolution of the flare with time, one has to note that it is highly dynamic on a typical time scale of 15 minutes, which expresses itself in all parameters: the overall flux (double peak), polarised flux, polarisation angle, and degree of polarisation.

\section{Context}

\begin{table*}
\centering
\begin{minipage}{14cm}
\caption{Properties of infrared flares observed since 2002. Observations were obtained in photometric (``phot''), polarimetric (``pol''), and spectroscopic (``spec'') modes. $\alpha$ is the colour index (defined as $\nu L_{\nu} \propto \nu^{\alpha}$), $p$ the degree of polarisation, and $q$ the angle of polarisation. Parameters marked ``--'' were not measured. Typical uncertainties are for fluxes: 1 mJy ($L'$ band: 3 mJy); times: 2 min;  $\alpha$: 1; $p$: 5\%; $q$: 10$^{\circ}$.}
\begin{tabular}{@{}rlllllllll@{}}

\hline
no. & epoch   & mode of     & band  & peak flux & duration & time scale of & $\alpha$ & $p$     & $q$	  \\
    &         & observation &       &           & of flare & sub-structure &	         &         & 	          \\
    & (years) &             &       &  (mJy)    & (min)    & (min)	      &  (1)     & (\%)    & ($^{\circ}$) \\
\hline
1   & 2002.66  & phot	& $L'$  & 30	    & $>$15    & --	       &  --	  & --      & --	   \\
2   & 2003.35  & phot	& $H$	& 16	    & 30       & --	       &  --	  & --      & --	   \\
3   & 2003.45  & phot	& $K_s$ & 13	    & 80       & 20	       &  --	  & --      & --	   \\
4   & 2003.45  & phot	& $K_s$ & 9	    & 85       & 17	       &  --	  & --      & --	   \\
5   & 2004.32  & phot	& $H$	& 9	    & $>$15    & --	       &  --	  & --      & --	   \\
6   & 2004.45  & pol	& $K_s$ & 5	    & 35       & --	       &  --	  & 20      & 80	   \\
7   & 2004.51  & phot	& $K_s$ & 8	    & $>$250?$^{\mathrm{a}}$ & 25   &  --     & --	 & --\\
8   & 2004.52  & phot	& $K_s$ & 3	    & 85       & 13	       &  --	  & --      & --	   \\
9   & 2004.54  & spec	& $K$	& 3	    & 60       & --	       &  -2.2    & --      & --	   \\
10  & 2004.54  & spec	  & $K$   & 3	      & 60	 & --		 &  -3.5    & --      & --	     \\
11  & 2005.27  & phot	  & $K_s$ & 3	      & $>$20	 & --		 &  --      & --      & --	     \\
12  & 2005.46  & spec	  & $K$   & 8	      & $>$150   & 20		 &  -3...+2$^{\mathrm{b}}$ & --      & --	\\
13  & 2005.57  & pol	  & $K_s$ & 8	      & 100	 & 20		 &  --      & 15      & 75	     \\
14  & 2006.40  & phot	  & $L'$  & 25        & 110	 & --		 &  --      & --      & --	     \\
15  & 2006.41  & pol	  & $K_s$ & 16        & $>$80	 & 15		 &  --      & 15...40$^{\mathrm{c}}$ & 80...10$^{\mathrm{c}}$ \\
16  & 2006.42  & phot	  & $L'$  & 23        & $>$150?$^{\mathrm{a}}$   & --	    &  --     & --	 & --		\\
\hline

\label{flaretable}
\end{tabular}

\begin{list}{}{}
\item[$^{\mathrm{a}}$] Very uncertain due to poor data quality.
\item[$^{\mathrm{b}}$] Variations within the same flare and correlation with source flux.
\item[$^{\mathrm{c}}$] Variations within the same flare with time.
\end{list}

\end{minipage}
\end{table*}

\begin{figure}
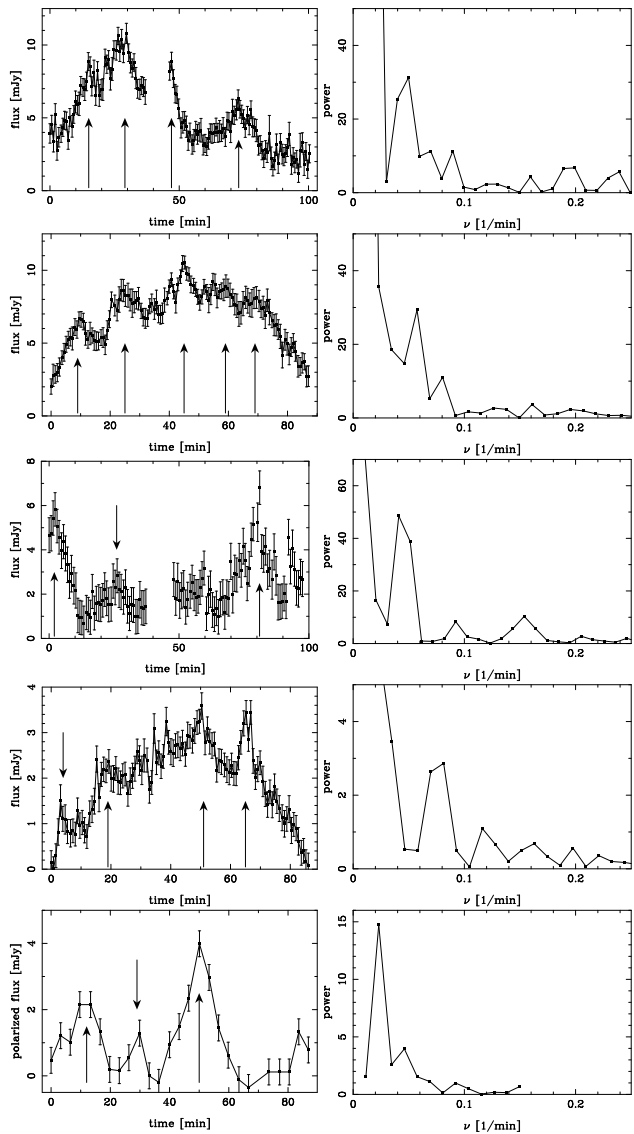


\centering

\includegraphics[height=4.1cm,angle=-90]{fig07a.eps}
\includegraphics[height=4.1cm,angle=-90]{fig07b.eps}

\includegraphics[height=4.1cm,angle=-90]{fig07c.eps}
\includegraphics[height=4.1cm,angle=-90]{fig07d.eps}

\includegraphics[height=4.1cm,angle=-90]{fig07e.eps}
\includegraphics[height=4.1cm,angle=-90]{fig07f.eps}

\includegraphics[height=4.1cm,angle=-90]{fig07g.eps}
\includegraphics[height=4.1cm,angle=-90]{fig07h.eps}

\includegraphics[height=4.1cm,angle=-90]{fig07i.eps}
\includegraphics[height=4.1cm,angle=-90]{fig07j.eps}

\caption{Lightcurves and power spectra illustrating the short-time sub-structure of SgrA* flares. Please note the different axes scales. The {\sl left} panels show the observed light curves. All data were obtained with NACO at the VLT in $K_s$ band. Local maxima in the overall flare shape are marked by arrows. Gaps in the light curves are due to intermediate sky observations. Times of observations are from top to bottom: June 15, 2003 (epoch 2003.45); June 16, 2003 (epoch 2003.45); July 6, 2004 (epoch 2004.51, first 100 min); July 8, 2004 (epoch 2004.52); July 29, 2005 (epoch 2005.57). The left bottom plot shows the polarised flux of the flare described by Eckart et al. (2006b). The {\sl right} panels show the corresponding Scargle periodograms of the light curves given in the left column. In case of the polarised flux, the periodogram is dominated by the $\sim$40-min distance between the two strongest peaks. All graphs show a secondary maximum corresponding to periods about 13--25 min, with accuracies about 2 min.}

\label{substructure}

\end{figure}

\begin{figure}
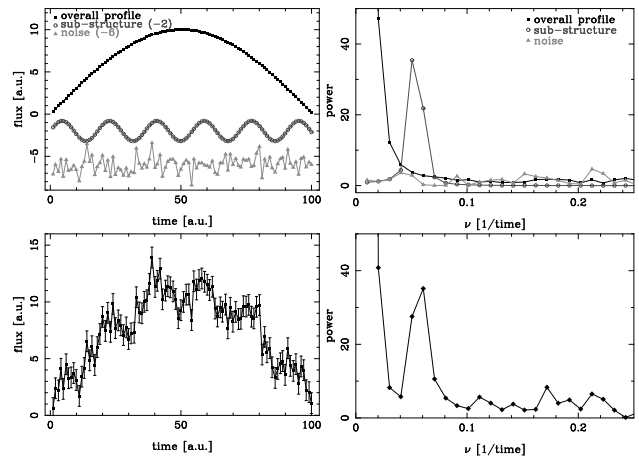


\centering

\includegraphics[height=4.1cm,angle=-90]{fig08a.eps}
\includegraphics[height=4.1cm,angle=-90]{fig08b.eps}

\includegraphics[height=4.1cm,angle=-90]{fig08c.eps}
\includegraphics[height=4.1cm,angle=-90]{fig08d.eps}

\caption{An artificial flare lightcurve as illustration of sub-structure signatures. All units are arbitary, but numerical values were selected thus that a comparison to Fig. \ref{substructure} is straight forward. {\sl Top left:} The additive components of the artificial flare: an overall profile (sine half-wave) with amplitude 10 and duration 100; a sine wave modulation with amplitude 1.2 and period 18; and Gaussian noise with $\sigma = 1$. {\sl Top right:} Scargle periodograms for each of the three components shown in the top left panel. {\sl Bottom left:} The synthetised flare lightcurve. This is the sum of the three components shown in the top left panel plus a constant background of height 1. To ease a comparison to real data, all data points are plotted with error bars with the length of the noise's $\sigma$. {\sl Bottom right:} Scargle periodogram of the light curve shown in the bottom left panel.}

\label{simflare}

\end{figure}

The SgrA* activity described above for the first time combines several separately observed properties on high significance levels: (1) strong, variable NIR activity with an overall duration of more than 80 minutes, (2) sub-structure on a time scale of 15 minutes, and (3) clear, variable polarisation.

These signatures allow a deeper understanding of the physical emission mechanisms of SgrA*. This is especially obvious in the long-term context of observed infrared, radio, and X-ray activity. For this reason the overall properties of SgrA* flares are discussed and compared here. 

Since 2002 we have observed the NIR emission of SgrA* using the ESO Very Large Telescope on Cerro Paranal, Chile. Photometric and polarimetric imaging data were collected with NACO in $H$, $K$, and $L$ bands (1.3 -- 4.1 $\mu$m; Genzel et al. \cite{genzel2003}, Trippe \cite{trippe2004}, Eckart et al. \cite{eckart2006a, eckart2006b}). 

Infrared spectra were obtained using SINFONI, a combination of the integral field spectrometer SPIFFI (Eisenhauer et al. \cite{eisenhauer2003a, eisenhauer2003b}) and the adaptive optics (AO) system MACAO (Bonnet et al. \cite{bonnet2003, bonnet2004}, at VLT-UT4. The data covered the $K$ band from 1.95 to 2.45 $\mu$m with a spectral resolution of $R = 4500$ (Eisenhauer et al. \cite{eisenhauer2005}, Gillessen et al. \cite{gillessen2006}). The properties of 16 observed infrared flares are given in Table \ref{flaretable}.

Additional to our work, infrared observations of SgrA* were obtained by Ghez et al. \cite{ghez2004, ghez2005} in $K$, $L$, and $M$ bands using the Keck II telescope on Hawaii. Emission in $L$ and $M$ bands observed with NACO was also reported by Cl\'enet et al. \cite{clenet2004, clenet2005}. Yusef-Zadeh et al. \cite{yusef2006} used the Hubble Space Telescope to monitor flares in the range 1.60--1.90$\mu$m. Recently, Krabbe et al. \cite{krabbe2006} collected spectroimaging data of a $K$ band flare using the integral field spectrometer OSIRIS at the Keck II telescope.

In radio wavelengths (sub-mm to cm) SgrA* was discovered by Balick \& Brown \cite{balick1974} and has since then been monitored extensively (recently e.g. Aitken et al. \cite{aitken2000}, Melia \& Falcke \cite{melia2001}, Bower et al. \cite{bower1999a, bower1999b, bower2003a, bower2005}, Miyazaki et al. \cite{miyazaki2004}, Marrone et al. \cite{marrone2006}, Macquart et al. \cite{macquart2006}) photometrically, spectroscopically, and polarimetrically using a large ensemble of telescope facilities.

In X-ray wavelengths (few keV) SgrA* flares have been observed photometrically and spectroscopically since 1999 using the NASA Chandra and the ESA XMM-Newton space telescopes (Baganoff et al. \cite{baganoff2001, baganoff2003}, Goldwurm et al. \cite{goldwurm2003}, Aschenbach et al. \cite{aschenbach2004}, B\'elanger et al. \cite{belanger2005, belanger2006}, Eckart et al. \cite{eckart2006a}).

This view over many seperate observations in several wavelength regimes allows some general statements:

\begin{itemize}

\item SgrA* emission is {\it flaring}. In NIR and X-ray wavelengths it is regularly detected in form of outbursts. In both bands flares correspond to an increase of flux by factors up to $\sim$10 from the background level within some ten minutes. The typical length of a flare is in the range of 1--3 hours. The flare event rate (i.e. the number of flares per time) is in the order of few events per day. For the 16 cases listed in Table \ref{flaretable} the flare rate is 2.5 events per day; including some flares covered by poor quality data increases this number to about $\sim$3.3 NIR events per day. In four cases NIR and X-ray flares were observed to be simulaneous within the available time resolutions (few minutes). Inspecting Table \ref{flaretable} shows (within the limits of low-number statistics) a general trend: flares are the more seldom, the more luminous they are. In contrast to this, changes in the radio flux are limited to variations of $<$50\% within hours to days.

The flaring character of SgrA* is illustrated by the lightcurves presented in Fig. \ref{fluxes}. Here the emission drops down to (and remains) zero (within the errors) in both total flux and polarised flux, after a phase of strong activity. Another example  is given in Fig. \ref{startingflare}. This figure presents an $H$ band flare observed in April 2004. In this case, after more than one hour of zero emission from the position of the black hole, strong emission raises up to 9~mJy within about 20 minutes. In both cases the observations are inconsistent with a permanent, variable NIR source. An equivalent behaviour could be observed at several other occasions in both NIR and X-ray bands. Thus the classification of SgrA* emission as ``flaring'' is justified.

\item SgrA* emission is {\it polarised}. Linear polarisation in the order of few to few ten per cent was detected in radio and NIR bands. In NIR, this polarisation is observed in flares as described in sections 2 and 3. For three NIR flares so far observed polarimetrically (flares 6, 13, and 15 in Table \ref{flaretable}) we found polarisation degrees of 15--20\% and angles of $\sim 80^{\circ}$ on sky at times of maximum fluxes. The polarisation fractions of flares 6 and 13 do not show significant variations with time. In contrast to the flare described in Sect. 3, there are no distinct peak/decay phases. Unfortunately, these statements are weakened by the larger relative errors caused by lower peak fluxes (5 and 8 mJy in contrast to 16 mJy) of the flares 6 and 13.

Concerning the observed polarisation angle, it is important to note that this angle was found repeatedly in three measurements covering a time span of two years. This strongly suggests that the geometry of the emission region is stable in time.

In comparison to this, the continuous radio flux was found to be polarised with $p\approx$ 2--8\% and $q\approx$ 135--$165^{\circ}$ (at $880\mu$m). The radio polarisation is variable (typically within the ranges given before) on time scales of few hours. Interestingly, Macquart et al. \cite{macquart2006}) find the intrinsic angle of polarisation to be about $165^{\circ}$. This would be close ($\sim25^{\circ}$) to our result for the decay phase of the flare described in Sect. 2 (modulo $180^{\circ}$). Such an agreement could indicate that at least in some phases of activity NIR and radio observations are tracing emission from the same region around SgrA*. 

Additional information has been found in the \emph{circularly} polarised radio emission. This was originally reported by Bower et al. \cite{bower1999b}. Based on a re-analysis of elder VLA data, Bower \cite{bower2003b} finds a constancy of the sign of the circular polarisation for about 20 years. This would -- again -- point towards a fixed B field orientation in the emission region.

\item SgrA* flares show a quasi-periodic {\it sub-structure} on time scales of minutes. Examples for this are given in Fig. \ref{substructure} (left column) where the lightcurves of five $K_s$ band flares observed from 2003 to 2005 are presented. The first four (from top) panels show fluxes vs. time, the fifth panel shows the polarised flux of the flare described by Eckart et al. \cite{eckart2006b}.

All flare lightcurves (especially panels 1, 2, 4) show characteristic structures: an overall profile (rise, maximum, decay) lasting about 1--2 hours is repeatedly modulated in cycles of 15--25 minutes.

In the right column of Fig. \ref{substructure} Scargle periodograms (Scargle \cite{scargle1982}) of the respective lightcurves are shown to visualise periodicities. The Scargle periodogram is defined as

\begin{equation}
P_{X}(\omega ) = \frac{1}{2}\cdot\left[\frac{\left(\sum_{j}^{}X_j\cos\omega t_j\right)^2}{\sum_{j}^{}\cos^2\omega t_j} + \frac{\left(\sum_{j}^{}X_j\sin\omega t_j\right)^2}{\sum_{j}^{}\sin^2\omega t_j}\right]
\end{equation}

Here $\omega$ is the angular frequency, $t_j$ the time of data point $j$, $X_j$ is the value measured at time $t_j$, and $P$ the power.

In order to illustrate the signatures presented in Fig. \ref{substructure}, Fig. \ref{simflare} gives a simple model in form of an artificial lightcurve. Please note that this model is an illustration only, not a simulation or reconstruction of a flare. The artificial flare is composed of four additive components (in arbitary units): (a) a sine half wave with length 100 and amplitude 10 as overall profile, (b) a sine wave with amplitude 1.2 and period 18 as periodic modulation, (c) random Gaussian noise with $\sigma=1$, and (d) a constant background of height 1. The sum of these four components forms the artificial flare. For each of the three non-constant components on the one hand and the final synthesised flare lightcurve on the other hand we computed the respective periodograms.

Comparing Figs. \ref{substructure} and \ref{simflare} allows to disentangle the features in the flare periodograms. These are: strong peaks at frequencies $\nu<0.02$ 1/min due to the overall flare profiles; secondary maxima at $\nu \approx$ 0.04--0.08 1/min due to the (quasi-)periodic substructure; and noise signals over the entire spectra.

Including this work, quasi-periodic signals in NIR flares have now been found in the range of 13--30 minutes. This sub-structure is generally quite weak -- indeed the ``double peak'' of the flare described in sections 2 and 3 is the strongest case seen so far -- and detected only in a part of all observed flares. This statement is true also for the X-ray activity of SgrA*, where quasi-periodic sub-structure with periods of about 5--22 min was reported for some flares.

\end{itemize}

\section{Discussion}

With all these elements at hand, we can start drawing fairly robust
conclusions on the nature of the flares. However, at this point we do not conclude that we can derive reliable quantitative parameters of SgrA*. Many parameters would still highly depend on the model assumptions made by the author. Therefore such a quantitative statement would necessarily suffer from over-simplification. The important physical facts or hints are better
obtained through qualitative discussion.

\subsection{Nature of the flares}

First of all, we know from imaging observations that flares occur always at the
same location, consistent within a few mas (a few 100 Schwarzschild radii,
$R_\mathrm{S}$) with the gravitational centre of the nuclear starcluster and the radio source Sgr~A* (Genzel et al. \cite{genzel2003}, Ghez et al. \cite{ghez2004}, Eisenhauer et al. \cite{eisenhauer2005}). But more importantly, the lightcurves of several NIR and X-ray flares observed so far show significant
variations on the timescale of 15~min. This demonstrates that the region
involved in these substructures is smaller than $\simeq10R_\mathrm{S}$.
Furthermore, the typical timescale in the lightcurves is consistent across these flares, ranging from 13 to 30~($\pm2$)~min.

This timescale is within the range of the innermost stable circular orbit
(ISCO; Bardeen et al. \cite{bardeen1974}) orbital periods allowed for Kerr black holes of $3$--$4\cdot 10^{6} \rm
M_{\odot}$ and various spin parameters. Many authors (Genzel et al. \cite{genzel2003}, Yuan et al. \cite{yuan2004}, Liu et al. \cite{liu2004}, Broderick \& Loeb \cite{broderick2006}, Paumard et al. \cite{paumard2006}) have studied the
possibility that the flare emission may actually come from matter orbiting the
BH close to the ISCO. The scatter in the observed periods is not a concern: in
the context of this ``orbiting blob scenario'', it would simply indicate that
the outbursts do not always occur exactly on the ISCO, but that a range of
orbital radii is allowed. Orbits \emph{inside} the ISCO are unstable. Since
flares last for more than one orbital period (typically more than four), we can assume that flares occur \emph{outside} this orbit. For this reason, the shortest period ever measured ($13\pm2$~min;
Fig.~\ref{substructure} and Table~\ref{flaretable}) sets a lower limit to the
spin parameter $a$ of the BH: using $M_{\rm SgrA*}=3.6\pm0.3\cdot 10^{6} \rm M_{\odot}$
(Eisenhauer et al.  \cite{eisenhauer2005}), this leads to
$a \geq 0.70\pm0.11$ (following Bardeen et al. \cite{bardeen1974}).

The presence of this quasi-periodic sub-structure imposes serious limits on alternative emission scenarii. \emph{Bow shock fronts} caused by stars moving through the accretion disc material (Nayakshin et al. \cite{nayakshin2003}) should not show such modulation.

In case of \emph{jet emission} (Falcke \& Markoff \cite{falcke2000}, Markoff et al. \cite{markoff2001}, Yuan et al. \cite{yuan2002}) such modulations would be imprinted on the jet if the jet nozzle was located in the accretion disc, orbiting the black hole. Indeed, the jet model by Falcke \& Markoff \cite{falcke2000} requires a nozzle with a radius of $\sim4~R_\mathrm{S}$ and a height of $\sim8~R_\mathrm{S}$. This extension would be small enough to allow for the observed short-time variability. On the other hand, Gillessen et al. \cite{gillessen2006} analyse the cooling time scales of orbiting hot spots and find a limit on the extension of the emission region of 0.3~$R_\mathrm{S}$. As this would be one order of magnitude smaller than the size of the model jet nozzle, the emission in the \emph{early} phases of a flare probably cannot be explained by a ``pure'' jet emission model. This does not exclude the presence of a jet but makes it unlikely that a jet is responsible for a significant part of the observed NIR emission. In the \emph{end} or \emph{decay} phase of a flare, when a substantial shearing and broadening of the emission region is expected from the hotspot model, a growing contribution from resp. evolution into a jet is thinkable; we will pay attention to this again later.

Additionally, the observations also do not a priori exclude the possibility of spiral \emph{density waves} propagating in the accretion disc. Those oscillations have been discussed in the context of stellar black hole systems in order to explain high-frequency (kHz range) QPOs (e.g. Kato \cite{kato2001}, P\'etri \cite{petri2006}, and references therein). A rotating two-arm structure could double the orbital time scale and thus loose the constraints on the BH spin. But those structures are expected to have life times which are shorter than the time of a single orbital revolution (Schnittman et al. \cite{schnittman2006}); this does not agree with the observations of flares lasting for several orbital periods (up to hours).

Our description might be somewhat challenged by X-ray observations for which
quasi-periodicities as short as 5~min have been claimed (Aschenbach et al.
\cite{aschenbach2004}, Aschenbach \cite{aschenbach2006}).  Using the dynamical
picture described above, such a short period would require a spin of about
$a=0.99$. Indeed Aschenbach et al. \cite{aschenbach2004} claim the
detection of several, resonant frequencies in the same flares. They interpret
this as a signature of oscillations in the accretion disc, leading to a spin
of $a=0.996$ and a mass of $M_{\rm SgrA*}=3.3\cdot 10^{6} \rm M_{\odot}$
(Aschenbach \cite{aschenbach2006}). However, B\'elanger et al.
\cite{belanger2006} find one periodicity of 22~min and no additional signals
in the same data sets. They developed a rigorous statistical method that excludes other (quasi-)periods being present to a statistically significant level in the data. This 22-min-period falls in the range of
the periods observed in NIR flares. From now on, we implicitly assume that
the flare emission comes from matter orbiting the BH.

The fact that the flare emission is polarised nicely confirms the synchrotron
radiation nature of the emitted light. This was already suspected from the overall
spectral energy distribution, NIR and X-ray colour indices, and the occasional
occurrence of simultaneous NIR and X-ray flares (Zylka \& Mezger \cite{zylka1988}, Zylka et al. \cite{zylka1992}, Baganoff et al. \cite{baganoff2001, baganoff2003}, Yuan et al. \cite{yuan2004}, Liu et al. \cite{liu2004}, Eckart et al. \cite{eckart2006a}). In this context, the polarisation parameter curves (Figs. \ref{fluxes}, \ref{params})
convey information about the geometry of the magnetic field. The remarkable
permanence of the polarisation parameters, in particular polarisation angle,
across three NIR flares observed over 2 years (Eckart et al. \cite{eckart2006b}, this work) indicates that the magnetic field geometry as well as the orbital plane remained the same for all three events.
This shows that the flaring material has enough time to settle in the BH's
equatorial plane before the occurrence of the flare, and speaks in favour of a
somewhat permanent accretion disc experiencing energetic events rather than
temporary structures building up randomly for each flare.

\subsection{Geometry of the system}

When comparing our data to the models by Broderick \& Loeb \cite{broderick2006}, the
non-detection of variations in the polarisation \emph{angle} during the peak phase
strongly suggests that the accretion disc is seen (within few degrees) edge-on. The non-detection of variations in the polarisation \emph{fraction} also speaks for an edge-on view of the disc. This
parameter is not exactly constant, but it would show only a short dip ($\sim$2~min) that our time
sampling would not allow detecting. Additionally, the constancy of the
polarisation fraction (about 15\% during the peak phase) speaks against the picture of (a) a dominating, slow component of the lightcurves (e.g. emission from the disc itself) on top of which (b) the periodic signal due to a second component is seen. Such a second component would be unlikely to be subject to the
same magnetic field as the bright spot itself. On the
contrary, the mechanism proposed by Paumard et al. \cite{paumard2006} by which this
slow component is due to shearing of the hotspot, evolving into a ring, fits
this observational result well.

There are basically two possible geometries for the magnetic field in the
orbiting spot scenario: poloidal (perpendicular to the orbital plane) and
toroidal (tangential to the orbit). The poloidal field is more natural in the
absence of matter or outside of the disc. On the other end, the field inside
the disc is most likely frozen and dragged by the matter. Due to shear, this naturally leads to a toroidal field (De Villiers et al. \cite{devilliers2003}, Broderick \& Loeb \cite{broderick2006}). A transition region above the disc and at its inner edge must exist, in which the magnetic field is somewhat
disorganised. This explains the fairly low observed polarisation fraction
($\sim$15\%; in a perfectly organised field, the polarisation fraction of
synchrotron emission is of order 75\%, Pacholczyk \cite{pachol1970}). The question remains which
component of the field is dominant; we will come back to this later.

The decay part of the flare reported here (sections 2, 3) shows a dramatic change in both
polarisation fraction (from 15\% to 40\%) and polarisation angle (from 80$^{\circ}$ to 10$^{\circ}$). It follows that the magnetic field seen by the electrons also changes dramatically. It becomes much more
organised, leading to an increased polarisation fraction, and rotates by
$\simeq70^\circ$. There are two options to explain this change: either the
field changes where the electrons are, or the electrons move to region with a
different field geometry. We will explore both possibilities below.

Let us first assume the flaring material remains on the orbital plane: in this
case, a change in the magnetic field could be due to the fact that the
accretion disc vanishes, letting the magnetic field relax into its matter-less
state, which is poloidal. This means that the field was mostly toroidal during
the peak phase.  The same conclusion is reached if the matter leaves the disc
from its inner edge, falling onto the BH.

The other possibility is that the material moves out of the accretion disc.
Since the flares are magnetically driven, it seems natural to assume that this
matter would follow the field lines, perhaps into a jet. Here again, the final
magnetic field is likely poloidal, hence the initial field is toroidal.

In these two schemes, the field is toroidal during the peak phase and poloidal
during the decay phase. We now assume that this is the case. A toroidal field
in the peak phase is yet another hint that the material has settled into a
disc and been able to drag the field before the occurrence of the flare.

The orientation of the magnetic field with respect to the
Galactic plane, which is located at $+27^{\circ}$, contains additional information. Indeed the peak phase polarisation angle is roughly perpendicular to the Galactic plane (to within $\sim$30 degrees), whereas the
decay phase polarisation angle is mostly in the Galactic plane (to within $\sim$10 degrees). We therefore see here an indication that the accretion disc of
Sgr~A* lies essentially in the plane of the Galaxy, and that its spin axis is
essentially aligned with that of the Galaxy. But as long as there are no stricter constraints on the polarised NIR emission from SgrA*, it is possible that future observations revise this picture.

\subsection{Proposed model}

Given together, we state that our data support the following model: SgrA* is a
fairly rapidly (perhaps maximally) rotating BH. Its spin axis is essentially
aligned with that of the Galaxy. It is surrounded by a somewhat permanent
accretion disc, with an inner edge close to the ISCO, in which the magnetic
field is toroidal. Outside of this disc, the field is poloidal. Occasionally,
shear will bend the magnetic field so much that a magnetic reconnection is
warranted. This is most likely to occur near the inner edge of the disc, where
shear is most effective. The magnetic reconnection heats a fraction of the
electrons to a hot temperature ($\simeq 10^{12}$ K). The region affected is
localised, smaller than the contraint imposed by cooling-time arguments in
Gillessen et al. (\cite{gillessen2006}): $R<0.3R_\mathrm{S}$. These electrons swirl in the
toroidal magnetic field and emit synchrotron emission. The emitting region
orbits the BH, giving raise to the periodic signal we observe. Shear as well
as magnetic forces make the region extend along the orbit. Since it spans only
a small range in distance from the BH, the shear is not extremely fast and
allows the periodic signal to be discernable for several periods.
Nevertheless, within a few orbital periods, the entire ISCO glows in
synchrotron emission, and this emission is responsible for the dominating,
slow part of the lightcurves (Paumard et al. \cite{paumard2006}). After some time, the
magnetic reconnection is over, removing the heating mechanism from the
picture. The electron population cools down, and at the same time extend
outside of the disc, perhaps into a jet. The dominating field then becomes
poloidal.

\section{Summary}

On May 31st, 2006, we observed a $K_s$ band flare from SgrA* which shows

\begin{itemize}

\item a high level of total flux, up to $\sim$16~mJy;

\item strong, variable polarisation, with $p=$15...40\%;

\item a polarisation angle between $\sim 80^{\circ}$ (during the peak phase) and $\sim 10^{\circ}$ (in the decay phase), swinging within about 15 minutes;

\item repeated sub-structure (double peak in total and polarised flux) on a time scale of 15~min.

\end{itemize}

Using this as well as information gathered during the last years from radio, NIR, and X-ray observations, we see strong indication that the flare emission in SgrA* is the synchrotron emission from material orbiting
the BH. We also find indication that some of this material eventually makes it
into a jet, reconciling the ``orbiting spot scenario'' tenants with the jet
hypothesis literature (Markoff et al. \cite{markoff2001}, Yuan et al. \cite{yuan2002}).

Finally, we might have observed the first pieces of evidence that a SMBH spin
axis is aligned with that of its host galaxy.

\section*{Acknowledgements}

Special thanks to N. Ageorges, ESO, for helpful discussions on NACO. We are grateful to the ESO instrument scientists and engineers who made possible this successful work. F.M. acknowledges support from the Alexander von Humboldt Foundation. We also would like to thank the anonymous reviewer whose comments helped to improve the quality of this article.

\label{lastpage}

\end{document}